\documentstyle{l-aa}
\input psfig 
\psfull

\def\be{\begin{equation}}
\def\ee{\end{equation}}
\def\ba{\begin{eqnarray}}
\def\ea{\end{eqnarray}}
\def\wisk#1 {\ifmmode{#1}\else{$#1$}\fi}
\def\ob    {\Omega_{b}}
\def\ol    {\lambda_{o}}
\def\oo    {\Omega_{o}}
\def\oc    {\Omega_{cdm}}
\def\etal   {{\sl et al.}~\rm}
\def\apj    {{\sl Ap.J.}\rm}

\def\chisq    {$\chi^{2}\:$}

\begin{document}
\thesaurus{
02  
(12.03.1;      
12.03.3;       
12.03.4)        
}
\title{Constraints on  $h$,  $\ob$ and $\ol$ from 
Cosmic Microwave Background Observations}
\author{ Charles H. Lineweaver \inst{1}
\and Domingos Barbosa  \inst{1,2}
\and Alain Blanchard  \inst{1}
\and James G. Bartlett \inst{1}
}
\institute{Observatoire de Strasbourg, 67000 Strasbourg, France. 
\and
Centro de Astrof\'{\i}sica da U.P., Rua do Campo Alegre 823, 4150 Porto, Portugal.}

\offprints{Charley Lineweaver, charley@cdsxb6.u-strasbg.fr}

\date{Received 1 August 1996/ Accepted 25 October 1996}
\maketitle

\begin{abstract}
In this paper we compare data to theory.
We use a compilation of the most recent cosmic microwave background (CMB) 
measurements to constrain Hubble's constant $h$, the baryon fraction $\ob$,  
and the cosmological constant $\ol$. We fit $h$-, $\ob$- and $\ol$-dependent
power spectra to the data.
The models we consider are flat cold dark matter (CDM) dominated
universes with flat ($n_{s}=1$) power spectra, thus the results obtained 
apply only to these models.
CMB observations can exclude more than 
half of the $h - \ob$ parameter space explored.
The CMB data favor low values of Hubble's constant; $h \approx 0.35$. 
Low values of $\ob$ are preferred ($\ob \sim 0.03$) but the \chisq 
minimum is shallow and we obtain $\ob < 0.28$.
A model with $h\approx 0.40$, $\ob \approx 0.15$ and  
$\oc \approx 0.85$ is permitted by constraints from the CMB data, 
BBN, cluster baryon fractions and the shape parameter $\Gamma$ derived 
from the mass density power spectra of galaxies and clusters. 

For flat-$\ol$ models, the CMB data, combined with BBN  constraints 
exclude most of the $h - \ol$ plane. 
Models with $\oo \approx  0.3$, $\ol \approx 0.7$ with $h \approx 0.75$ 
are fully consistent with the CMB data but are excluded by the strict new 
$q_{o}$ limits from supernovae (Perlmutter \etal 1997).
A combination of CMB data goodness-of-fit statistics, BBN and supernovae 
constraints in the $h-\ol$ plane,  limits Hubble's constant to the 
interval $0.23 < h < 0.72$.

\end{abstract}

\section{Introduction}
\label{sec:intro}

A new technique is coming on-line producing a 
small revolution in our ability to evaluate cosmological models 
(Bond \etal 1994).
Measurements of fluctuations in the cosmic microwave background
(CMB) over a large range of angular scales have become sensitive
enough to distinguish one model from another.
This technique is truly cosmological and independent of previous methods.
It probes scales much larger and
times much earlier ($z \approx 1000$) than more traditional techniques 
which rely on supernovae, galaxies, quasars and other low-redshift objects.

Our ignorance of $h$, $\ob$ and $\ol$ is large
and there are many inconsistent observational results.
For example, after 60 years of effort the favorite Hubble constants of 
respected cosmologists can differ by more than a factor of 2 
($0.40 \la h \la 0.90$) with error bars small compared to this interval.
Thus, information on Hubble's constant and other cosmological parameters from
the independent and  very high redshift CMB data is important.
Over the next decade observations
at small angular scales have the potential to determine many important 
cosmological parameters to the $\sim 1\%$ level (Jungman \etal 1996). 
We expect CMB measurements to tell us 
the ultimate fate of the Universe ($\oo$),
what the Universe is made of ($\ob$, $\oc$) and 
the age and size of the Universe ($h$) with unprecedented precision.

The COBE detection of temperature fluctuations in the CMB
(Smoot \etal 1992, Bennett \etal 1996) has constrained
the amplitude and slope of the power spectrum at large angular scales.
Scott \etal (1995), using a synthesis of the most current CMB measurements at 
that time, demonstated the existence of the predicted acoustic peak.
Using compilations of CMB measurements,
Ratra \etal (1995), Ganga \etal(1996), G\'orski \etal(1996), White \etal(1996) 
have compared groups of favored models to the data and obtained interesting 
constraints on $\Omega_{o}$, $n$ and the normalizations of various flat and open
CDM models.

The most recent ground-based and balloon-borne experiments 
(Netterfield \etal 1997, Scott \etal 1996, Platt \etal 1996, Tanaka \etal 1996)
are providing increasingly accurate CMB fluctuation measurements on 
small angular scales.
In this work we take advantage of these new measurements and of a fast Boltzmann 
code (Seljak \& Zaldarriaga 1996) to make a detailed exploration of 
two large regions of parameter space: the $h - \ob$ and $h - \ol$ planes. 
Such an approach allows us to place independent CMB-derived constraints on 
$h$, $\ob$ and $\ol$ based on goodness-of-fit statistics.
 
Our results are valid only in the context of the models we have considered;
we assume COBE normalized, Gaussian, adiabatic initial
conditions in spatially flat universes ($k=0$, $\oo + \ol =1$) 
with Harrison-Zel'dovich ($n_{s}=1$) temperature fluctuations.
For Gaussian fluctuations the power spectra, $C_{\ell}$, uniquely
specify the models.
Our limits include estimates of the uncertainty due to 
the COBE normalization uncertainty
as well as the Saskatoon absolute calibration uncertainty.
We use $h=H_{o}/100$ km s$^{-1}$ Mpc$^{-1}$.

Our CMB-derived  constraints on $h$, $\ob$ and $\ol$ are independent of BBN and other
cosmological tests.
Where the CMB constraints are not as tight as other methods, 
the existence of overlapping regions of allowed parameter space is an 
important consistency check for both;
unknown systematic errors can be uncovered by such a comparison.

In Section 2 we describe the \chisq calculation.
In Section 3 we provide an overview of the physics of
fluctuations and the power spectrum models used in the fit.
In Sections 4 and  5  we discuss our results for the $h-\ob$ and 
$h -\ol$ planes respectively and  combine them with a variety of other 
cosmological constraints.
In Section 6 we summarize and discuss our results.

\section{Data Analysis}
\label{sec:analysis}

\subsection{The Recipe}

\noindent $\bullet$ assemble all available CMB experimental results with the 
corresponding window functions

\noindent $\bullet$ determine the region of parameter space one would like
to explore (we choose the  $h- \ob$ and $h$ - $\ol$ planes for 
spatially flat models)
  
\noindent $\bullet$ 
use a fast Boltzmann code (Seljak \& Zaldarriaga 1996)
 to obtain the power spectra for a matrix of models 
covering the desired region of parameter space

\noindent $\bullet$ convolve each model with the experimental window functions
and fit the result to the data by producing a \chisq surface over
the chosen parameter space

\noindent $\bullet$ compare the results with other cosmological constraints

\subsection{The Equations}
\begin{figure}[htbp]
\centerline{\psfig{figure=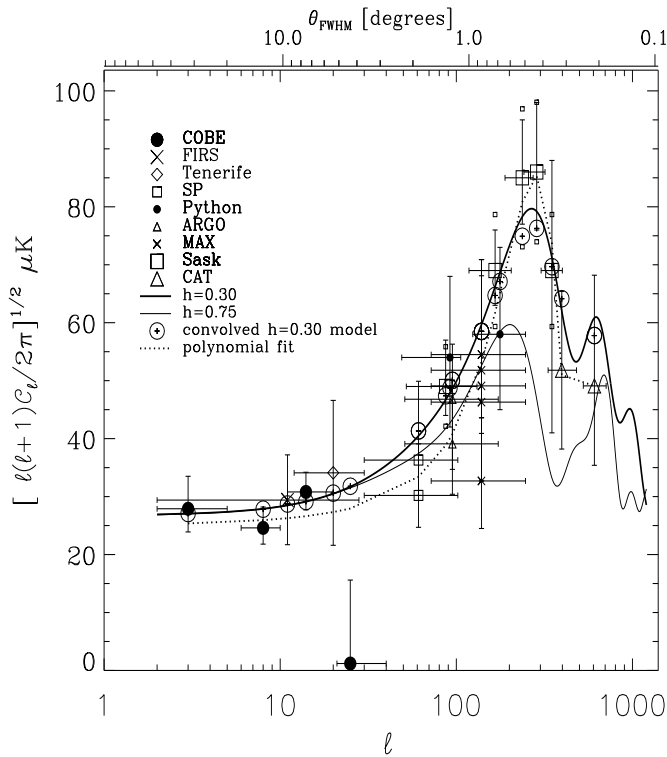,height=10.0cm,width=9.cm,bbllx=10pt,bblly=120pt,bburx=594pt,bbury=690pt}}
%
\caption{ 
The power spectrum of observed CMB temperature fluctuations
(from Table 1) as a function of angular scale.
The FWHM of the window functions are indicated by the horizontal
lines at each data point.
Two representative models are also plotted;
both have $\Omega_b = 0.05$. The differences between the 
convolved $h=0.3$ model (large circles) and the measurements 
is used to calculate a single point of the \chisq surface shown in 
Figure \protect\ref{fig:conv_chi}.
The small boxes above and below the Saskatoon points indicate the
$\pm 14\%$ absolute calibration uncertainty shared by these 5 points.
The dotted line is a 5th order polynomial fit to the data.
On the upper axis, 
$\theta_{FWHM} = (32\: ln\:2)^{1/2} sin^{-1}(\frac{1}{2\ell + 1})$
(Bond 1995).
}
\label{fig:data}
\end{figure}

We calculate the \chisq surface over a matrix of models 

\be
\chi^{2}(i,j)= \sum_{N=1}^{N_{exp}}
\left[ \frac{\delta T_{\ell_{eff}}^{data}(N)-\delta T_{\ell_{eff}}^{model}(N,i,j) }
{\sigma^{data}(N)}\right]^{2},
\ee
and use it to indicate the regions of parameter space preferred by the data.
The sum is over the CMB detections plotted in 
Figure \ref{fig:data} and listed in Table 1 in the form of 
flat band power: $\delta T_{\ell_{eff}}^{data}(N)$
with the corresponding errors $\sigma^{data}(N)$.
The models are indexed in the 2-D parameter space by $i$ and $j$.
The band power estimates of the data and models are defined respectively as
\ba
\nonumber
\delta T_{\ell_{eff}}^{data}(N)\:\:&=&\:\: 
\frac{\delta T^{obs}_{rms}(N)}{\sqrt{I(W_{\ell}(N))}}\\
& =&
\left[\frac{1}{I(W_{\ell}(N))}\sum_{\ell=2}^{\ell_{max}}
\frac{(2\ell + 1)}{4\pi}C_{\ell}(real) W_{\ell}(N)\right]^{1/2}
\ea

\ba
\nonumber
\delta T_{\ell_{eff}}^{model}(N,i,j)&=&
\frac{\delta T^{model}_{rms}(N,i,j)}{\sqrt{I(W_{\ell}(N))}}\\
 &=&
\left[\frac{1}{I(W_{\ell}(N))} \sum_{\ell=2}^{\ell_{max}}
\frac{(2\ell + 1)}{4\pi}C_{\ell}(i,j) W_{\ell}(N)\right]^{1/2},
\ea
where $\delta T^{obs}_{rms}(N)$ is the rms temperature fluctuation 
observed by the $Nth$ experiment,
$W_{\ell}(N)$ are the experiment-specific window functions
(White \& Srednicki 1995) and the
deconvolving factors $I(W_{\ell}(N))$ are the logarithmic integrals 
of the $W_{\ell}$ (Bond 1995) defined as 
\be
I(W_{\ell}) = 2\pi \sum_{\ell=2}^{\ell_{max}}
\frac{(2\ell+ 1)}{4\pi}\frac{W_{\ell}}{\ell(\ell+1)},
\ee
where $\ell_{max}=1200$ since for all $N$, $W_{\ell}(N) \approx 0$ 
for $\ell > 1200$.  
The angular scale probed by the $N$th experiment is
\be
\ell_{eff}(N) = \frac{I(\ell\: W_{\ell}(N))}{I(W_{\ell}(N))}.
\ee

In equation (2),
the convolution of the unknown real power spectrum of the CMB sky, $C_{\ell}(real)$, 
with $W_{\ell}$ is the observed temperature rms, $\delta T^{obs}_{rms}$.
One cannot compare the rms results from different experiments with 
each other unless the influence of the window function 
has been removed, i.e., deconvolved. 
The division by $I(W_{\ell})$ is this deconvolution 
and works optimally when
$\ell(\ell + 1)C_{\ell}(real)$ is a constant across the range of
$\ell$ sampled by the experiment. 
Equation (4) has been written suggestively to clarify this deconvolution.
Notice that the models are treated like the real sky:
first they are convolved with the window function and then the division
by $I(W_{\ell})$ deconvolves the window function.
Thus our model points are ``convolved-deconvolved'' power spectra.
Setting  $C_{\ell}= a/\ell(\ell + 1)$ in equation (2) or (3) yields
$\delta T_{\ell_{eff}} = (a/2\pi)^{1/2} = [\ell(\ell + 1)C_{\ell}/2\pi]^{1/2}$ 
which is the origin of the units of the y-axis of Figure \ref{fig:data}  and why 
it is reasonable to plot the input model and the flat band power estimates 
on the same plot.

\begin{figure}[htbp]
5579F2
\centerline{\psfig{figure=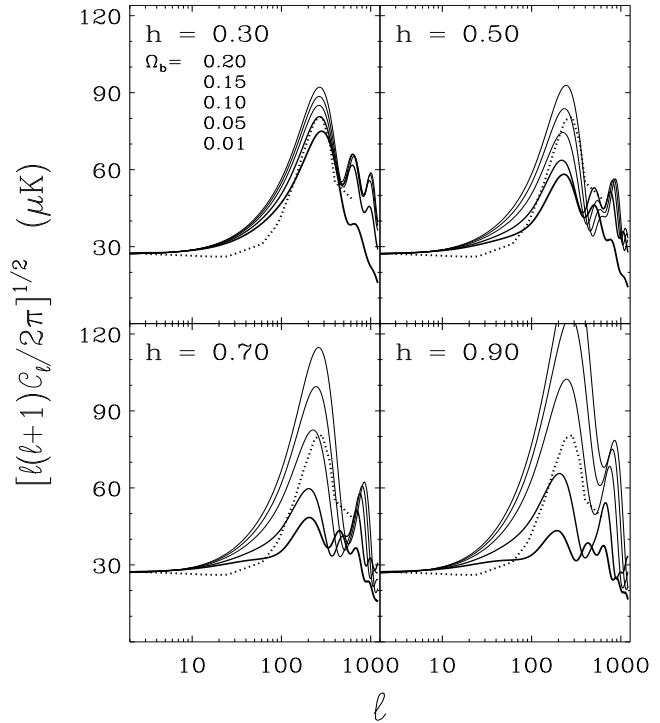,height=10cm,width=\hsize,bbllx=20pt,bblly=50pt,bburx=547pt,bbury=722pt}}
%
\caption{
Representative power spectra showing the $h$- and $\ob$-dependence.
In each panel $h$ is fixed while $\ob$ takes on the values indicated.
The largest values of $\ob$ have the largest Doppler peaks.
Notice that as  $h$ increases, $A_{peak}$ increases for large $\ob$ but
decreases for small $\ob$; thus at high $h$ the peak height is an excellent baryometer.
The dotted line is the polynomial fit to the data in
Figure \protect\ref{fig:data} and is the same in all panels.
All models are spatially flat ($k=0$, $\ol = 0$), Harrison Zel'dovich ($n_{s} = 1$)  
normalized to the COBE 4-year results at $\ell=6$.
} 
\label{fig:conv_cl}
\end{figure}

Using the data and equation (3) in equation (1) yields a \chisq value for 
every point in the matrix of models.
Figure \ref{fig:data} is a picture of the ingredients of our \chisq calculation.
The CMB measurements from Table 1 are plotted. 
The thick solid line is an $h=0.30$ model with $\Omega_b = 0.05$.
The large open circles are the convolution of this $h=0.30$ model with the
experimental window functions. The difference between the resulting
``convolved-deconvolved'' points and the measurements is used to calculate 
the \chisq values.
By comparing the data with these ``convolved-deconvolved'' 
points rather than with the direct $[\ell(\ell + 1)C_{\ell}/2\pi]^{1/2}$ 
values of the original $h=0.30$ model,  we are accounting for the 
experimental window function even in the regions of sharp
peaks and valleys. 
The discrepancy between the open circles
and the original $h=0.30$ model (solid line) is a measure of the necessity 
of this convolution-deconvolution procedure; it is unnecessary
except in the sharp peaks and valleys.
The dotted line is a fifth order polynomial fit to the data points
and is used in Figures \ref{fig:conv_cl} and \ref{fig:hlam_cl} to represent 
the data.
Notice that the amplitude of the primary acoustic peak in the data is
$A_{peak} \sim 80 \: \mu$K.

Most experimental results contain slightly asymmetric error bars and, most commonly, 
the upper error bar is larger.
For a given data point, this asymmetry constrains models lower than the 
data point a bit more rigidly than it does models which are higher than 
the data point.
To approximate this asymmetry for the \chisq calculation,  we toggle
the error bar depending on whether the model is above or below the data point. 

\section{CMB Power Spectra}
\label{sec:power}
\subsection{General Features}

Figures \ref{fig:conv_cl} and \ref{fig:hlam_cl} are samples from the matrices 
of power spectra 
covering the parameter space explored.
They show the influence of $h$, $\ob$ and $\ol$
on the angular power spectrum of CMB fluctuations.
Generic features are: 

\noindent $\bullet$ a flat Sachs-Wolfe plateau for $\ell \la 30$ due to the 
gravitational potentials of superhorizon-sized density fluctuations.

\noindent $\bullet$ a primary acoustic peak at $\ell_{peak} \sim 250$
accompanied by secondary peaks and valleys
for $\ell \ga 2\:\ell_{peak}$ 

\noindent $\bullet$ a  cut-off at $\ell \ga 1000$ due to averaging along the line of sight as the
photons traverse the finite thickness of the surface of last scattering

The physics of the primary acoustic peak is the most relevant
for our purposes since this peak dominates the fits.
At sub-degree angular scales acoustic oscillations of the baryon--photon 
fluid at recombination produce peaks and valleys in the CMB power spectrum.
We may understand their general features by employing
the driven  oscillator interpretation of Hu (1995) and
Hu \& Sugiyama (1995a). The waves are set up by the pressure of the 
photons, which are tightly coupled to the baryons and subject
to the driving force of the gravitational potential.  Ignoring the
neutrinos, there are three particle species present at recombination 
(baryons, photons and dark matter) which determine two fundamental ratios:
the baryon--to--photon ratio, varying as $\ob h^2$, 
and the dark matter--to--photon ratio, varying as $\oc h^2$.
We expect the physics at this epoch to be largely invariant to changes
which leave these quantities unaltered, i.e., we expect $\ob h^{2}$
and $\oc h^{2}$  to control the shape of these oscillations and in
particular the height and location of the primary peak
$A_{peak}$, $\ell_{peak}$.
\begin{figure}[htbp]
\centerline{\psfig{figure=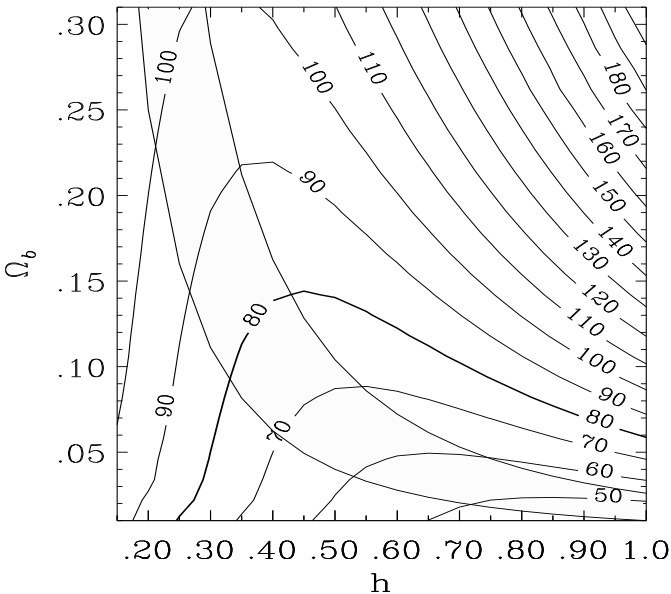,height=8cm,width=\hsize,bbllx=0pt,bblly=12pt,bburx=538pt,bbury=608pt}}
%
\caption{
Contours of $A_{peak}\: [\mu$K].
Although no data were used to make this plot, the $80\: \mu$K contour is thick
to indicate the approximate peak height of the data.
For large $h$, the strong $\ob$-dependence of the peak heights
is responsible for the narrow spacing between contours and for the 
narrow range of $\ob$ permitted by the data 
(see Figure \protect\ref{fig:conv_chi}).
As $h$ decreases, $A_{peak}$ loses its $\ob$-dependence
thus  permitting a wider range of $\ob$.
The grey band is the region favored by big bang nucleosynthesis: 
$0.010 < \ob h^{2} < 0.026$ (e.g. Copi \etal 1995, Tytler \& Burles 1997).
All models in this plot have $\oo = \ob + \oc  =1$ and  $\ol = 0$.
}
\label{fig:conv_a}
\end{figure}
\begin{figure}[htbp]
\centerline{\psfig{figure=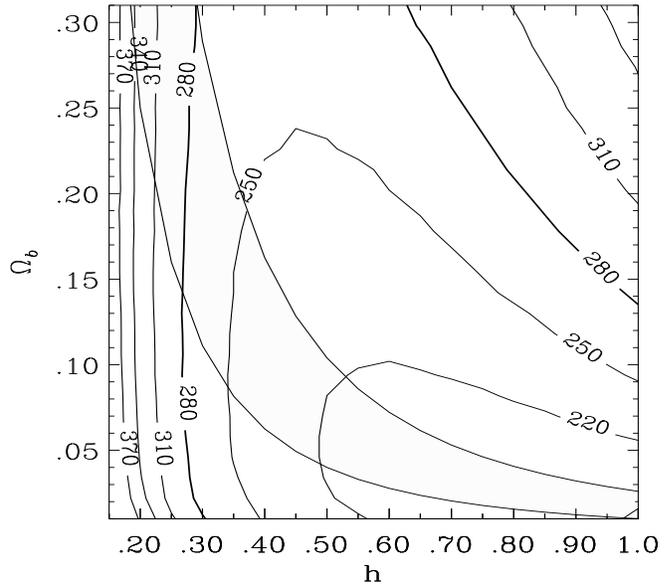,height=8cm,width=\hsize,bbllx=0pt,bblly=12pt,bburx=538pt,bbury=608pt}}
%
\caption{
Contours of $\ell_{peak}$ for the same models as in 
Figure \protect\ref{fig:conv_a}.
The thick 280 contour is indicative of the $\ell_{peak}$ favored by the data.
In the lower right, $\ell_{peak}$ is too small to suit the data.
} 
\label{fig:conv_l}
\end{figure}

\subsection{Effects of $\ob h^{2}$ and $\oc h^{2}$}
\subsubsection{Peak Height}
Figures \ref{fig:conv_cl}, \ref{fig:conv_a}, \ref{fig:conv_l}
are a triptych; they show three different ways of looking at
the $h - \ob$ models we have explored.
Figure \ref{fig:conv_cl} shows the entire power spectra but only for 
a sparse sample of parameter space.
Each panel shows 5 samples from a vertical strip
of Figures \ref{fig:conv_a} and  \ref{fig:conv_l}.
Figures \ref{fig:conv_a} and \ref{fig:conv_l}
fully sample the matrix of models but with only one-parameter
characterizations of the spectra.

It is important to distinguish aspects of the \chisq surface
produced by the underlying models from the aspects produced by the
data.
The data currently available are in the range $ 2 \la \ell_{eff} \la 600$
and are insensitive to most of the fine details 
in the models seen at  $\ell \ga 2\;\ell_{peak}$. 
Since the Doppler peak of the spectra is the most prominent feature, 
its amplitude $A_{peak}$ is the simplest 1-parameter characterization of the the relative 
topology of the parameter space and is an excellent tracer of what the 
\chisq contours will look like.

Figure \ref{fig:conv_a} is a  contour map of $A_{peak}$ created by 
plotting the maximum amplitude of each model power spectrum.
Although no data are involved,  the $A_{peak}=80\;\mu$K contour is thick to
indicate the approximate peak height of the data in Figure \ref{fig:data}.
If $A_{peak}$ were the only important feature relevant for the fit, the \chisq minimum 
would straddle the $A_{peak} \sim 80\: \mu$K contour 
and be contained by boundaries parallel to these peak contours.
This is almost the case 
(see Figure \ref{fig:conv_chi}).

In Figure \ref{fig:conv_a} one can see that for a given $h$,
$A_{peak}$ increases monotonically with $\ob$:
$A_{peak}$ is low at the bottom of the plot and high at the top.
The grey band marks the BBN region bounded by contours of
$\ob h^{2}=$ constant.
In the upper right, higher values of $\Omega_b h^2$ lead to larger Doppler 
peaks due to the enhanced compression caused by a larger effective mass 
(more baryons per photon) of the oscillating fluid.

If we move along the BBN region from right to left, $A_{peak}$ increases.
Since $\ob h^{2} = $ constant in this region, we are seeing the influence
of $\oc h^{2}$. As $h$ goes down, $\oc h^{2}$ goes down, and therefore
$A_{peak}$ goes up.
This can be understood more physically.
Lowering $h$ increases the fraction of the total energy density 
locked up in the photons.  As this component cannot grow in
amplitude (it is oscillating), it retards the potential evolution
with respect to purely matter--dominated fluid -- in other words,
it causes a {\it decay} in the gravitational potential.  An oscillator
driven by a decaying force will actually obtain larger amplitudes
around its zero point than one subject to a constant force, assuming
the same initial conditions.  At fixed $\Omega_b h^2$, this leads
to a larger primary acoustic peak.  

Contours parallel to the BBN  region with $A_{peak}$ increasing
to the upper right, are imposed by $\ob h^{2}$.
Nearly vertical contours are imposed by $\oc h^{2}$  
($\oc \approx \oo =1$)
with $A_{peak}$ increasing to the left.
In summary, 
$\ob h^{2}$ and $\oc h^{2}$ produce competing effects; when
$\ob h^{2}\Uparrow$ $A_{peak}\Uparrow$ and when
$\oc h^{2}\Uparrow$ $A_{peak}\Downarrow$.
The two effects combined explain the shape of the 
contours in Figure \ref{fig:conv_a}. 

\begin{figure}[htbp]
\centerline{\psfig{figure=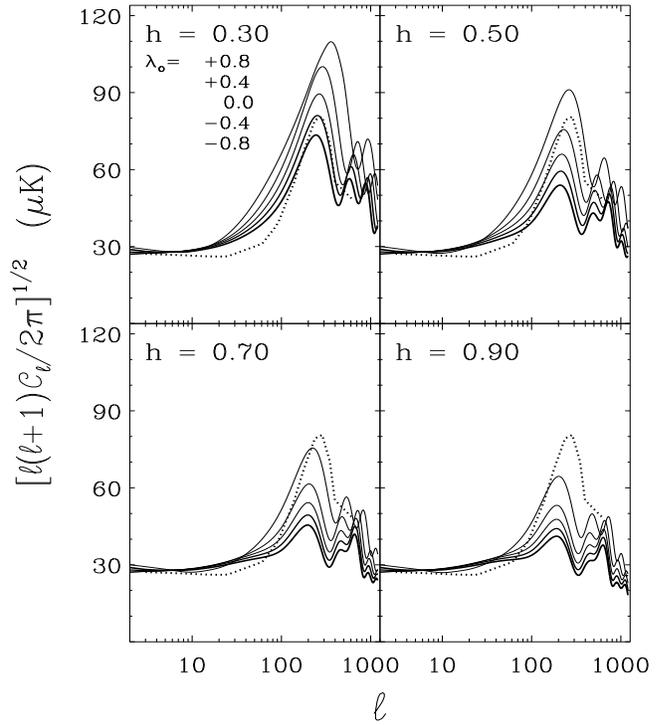,height=10cm,width=\hsize,bbllx=20pt,bblly=50pt,bburx=547pt,bbury=722pt}}
%
\caption{
Representative power spectra showing the $h$- and $\ol$-dependence.
In each panel $h$ is fixed while $\ol$ takes on the values indicated.
The largest values of $\ol$ have the largest Doppler peaks. 
This is predominantly the effect of $\oc h^{2}$ decreasing since the 
baryon-to-photon ratio has been fixed: $\ob h^{2}= 0.015$ and 
($\oc + \ob +\ol = 1$). 
Notice that as $h$ increases, $A_{peak}$ decreases and $\ell_{peak}$ shifts to 
larger scales.
The dotted line is the polynomial fit to the data in
Figure \protect\ref{fig:data} and is the same in all panels.
All models are spatially flat ($k=0$),  Harrison Zel'dovich ($n_{s} = 1$)  and 
normalized to the COBE 4-year results at $\ell=6$.
}
\label{fig:hlam_cl}
\end{figure}

\begin{figure}[htbp]
\centerline{\psfig{figure=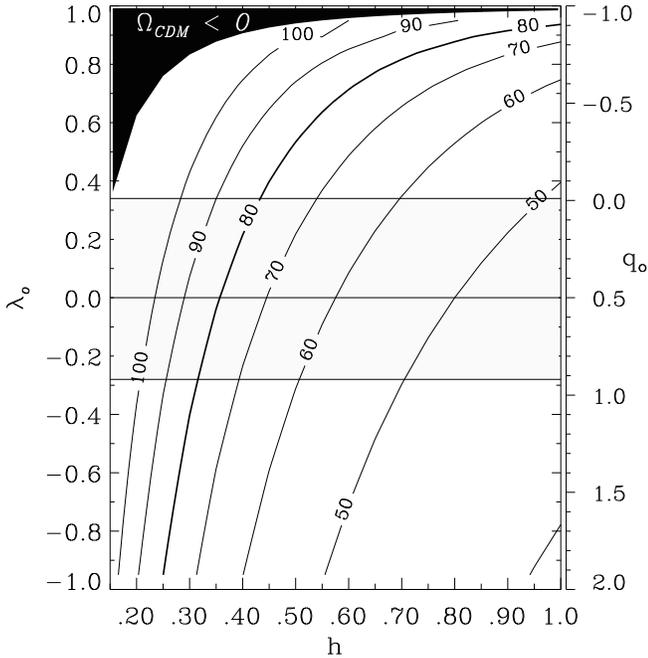,height=9cm,width=\hsize,bbllx=18pt,bblly=40pt,bburx=594pt,bbury=630pt}}
%
\caption{
Contours of $A_{peak}\: [\mu$K] with $\ob h^{2} = 0.015$.
As in Figure \protect\ref{fig:conv_a} no data were used to make this plot. 
The $80\: \mu$K contour is thick to indicate the approximate peak height of 
the data.
For a given $h$, as $\ol$ increases, $A_{peak}$ increases.
The grey area marks the permitted area ($1 \sigma$) from new supernovae results 
(Perlmutter \etal 1997). All models in this plot have $\oo + \ol = 1$.
} 
\label{fig:hlam_a}
\end{figure}

In Figures \ref{fig:hlam_cl} and \ref{fig:hlam_a} we see the effect of varying 
$\ol$ and $h$ with $\ob h^{2}$ held fixed.
Since $\ob h^{2}$ is fixed, for any given $\ol = constant$,
when $h \Uparrow \oc h^{2} \Uparrow$ and thus $A_{peak} \Downarrow$.
Also since $\ob + \oc + \ol = 1$, for any given $h$
when $\ol \Uparrow$ we have $\oc h^{2} \Downarrow$ thus $A_{peak}\Uparrow$.
We can see both of these effects in the $A_{peak}$ contours of 
Figure \ref{fig:hlam_a}.
As in Figure \ref{fig:conv_a} no data were used to make this plot and 
the $80\: \mu$K contour is thick to indicate the approximate peak height 
of the data.
%
\begin{figure}[htbp]
\centerline{\psfig{figure=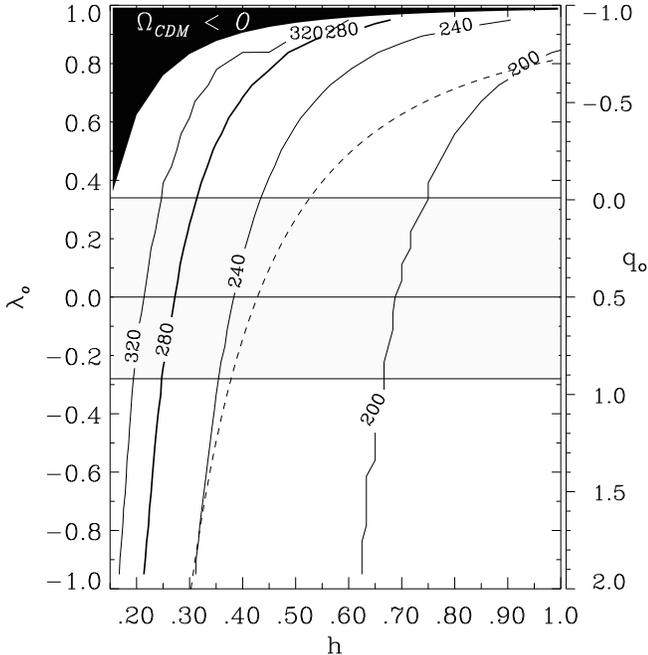,height=9cm,width=\hsize,bbllx=18pt,bblly=40pt,bburx=594pt,bbury=630pt}}
%
\caption{
Contours of $\ell_{peak}$ for the same models as in Figure
\protect\ref{fig:hlam_a}.
Although no data were used to make this plot, the thick 280 contour is in 
the center of the region favored by the data.
The dashed line is discussed in Section 3.2.2.
}
\label{fig:hlam_l}
\end{figure}

\subsubsection{Peak Position}
Hu \& Sugiyama (1995b) identify three factors which, when increased,
increase $\ell_{peak}$: $\ob h^{2}$, $\ol$ and $(1-\oo-\ol)$.
And one factor which, when increased, decreases $\ell_{peak}$, 
$\oo h^{2}$.
For all our models $1-\oo-\ol= 0$, so this focusing effect plays no
role here.
The effect of $\ob h^{2}$ can be understood using its relationship to the
sound speed at recombination:
$c_{s} = (1+a\ob h^{2})^{-1/2}$ where $a$ is a constant, thus
if $\ob h^{2} \Uparrow$ $c_{s}\Downarrow$. 
The size of the fluctuations is controlled
by the sound horizon at decoupling, thus $\theta_{peak} \propto c_{s}$.
If $c_{s} \Downarrow$ then   $\theta_{peak}\Downarrow$ and thus
$\ell_{peak} \Uparrow$.
We can see this clearly in the upper right of Figure \ref{fig:conv_l}.
$\ob h^{2}$ and $\oc h^{2}$ have opposite effects on the peak amplitude and
location. 
When $\ob h^{2} \Uparrow \ell_{peak}\Uparrow$ and $A_{peak} \Uparrow$.
When $\oc h^{2} \Uparrow \ell_{peak}\Downarrow$ and $A_{peak} \Downarrow$.
 
The location of the peak plays an important role in preferring the low 
$h$ region of iso-$A_{peak}$ contours.
The CMB data have $\ell_{peak} \sim 280$ which corresponds to an angular 
scale of $\sim 0^{\circ}\!.5$. 
In the lower right of Figures \ref{fig:conv_l} and \ref{fig:hlam_l}, 
$\ell_{peak} \sim 210$ (substantially lower than 280). Thus the data
disfavor these high $h$ models.

There is an interesting apparent inconsistency in 
Figure \ref{fig:hlam_l}. In the entire figure $\ob h^{2} = 0.015= constant$.
Along the dashed line  $\oo h^{2} = \ob h^{2} + \oc h^{2} = 0.015 + (0.41)^{2}= constant$.
Thus as we follow the dashed line up, $\ol$ increases and we should see
$\ell_{peak}$ increasing since, as we mentioned earlier
if $\ol \Uparrow$ then $\ell_{peak}\Uparrow$.
This is not the case however; as we follow the dashed line upwards, $\ell_{peak}$ decreases.
A scaling argument can clarify this.
The angle $\theta$ subtended by an object of size $ct$ at an angular distance $d_{ang}$ 
is $\theta \sim ct/d_{ang}$. Thus,
\be
\ell_{peak} \sim \frac{1}{\theta_{peak}} 
            \sim \frac{d_{ang}(h, \oo, \ol)}{c\;t_{rec}}
            \sim \frac{h^{-1}f(\oo,\ol)}{(\oo h^{2})^{-1/2}}
            \sim \frac{f(\ol)}{h}
\ee
where $\theta_{peak}$ is the angular scale of the Doppler peak.
The physical scale of the peak oscillations 
is some fixed fraction of the horizon $\propto c\;t_{rec}$.
The time of recombination scales as $(\oo h^{2})^{-1/2}$ 
(which is constant along the dashed line) and the
angular distance $d_{ang}= h^{-1} f(\oo,\ol)$.
Since in our flat models $\oo = 1-\ol$, $f(\oo, \ol) \rightarrow f(\ol)$
where 
\be
f(\ol) = \int_{0}^{z_{rec}} 
\frac{dz}{\left[ (1+z)^{3}- \ol(1+z)^{3}\right]^{1/2}}.
\ee
which, when inserted into equation (6) is an expression of the monotonic 
relation, when $\ol \Uparrow$ then $\ell_{peak}\Uparrow$.
However, for flat $\ol$ models with
$\ob h^{2}$ and $\oc h^{2}$ fixed, one cannot change $\ol$ without 
changing $h$. So, in the case we are considering equation (6) is telling us that
when $\frac{f(\ol)}{h}\Uparrow \ell_{peak} \Uparrow$.
The $h$ scaling can be understood as the effect of larger universes:
a given physical size at a larger distance subtends a smaller angle.

Following the pioneering work of Hu \& Sugiyama (1995a, 1995b), in this section  
we have presented contours of the solutions of the 
Boltzmann equation and we have discussed how the behavior of 
$A_{peak}$ and $\ell_{peak}$ can be explained 
in terms of $\ob h^{2}$ and $\oc h^{2}$. 
These contour plots are particularly relevant for
the next section where we describe the regions of solution
space preferred by the CMB data.

\section{ $h - \ob$ Results and Discussion}
\begin{figure}[htbp]
\centerline{\psfig{figure=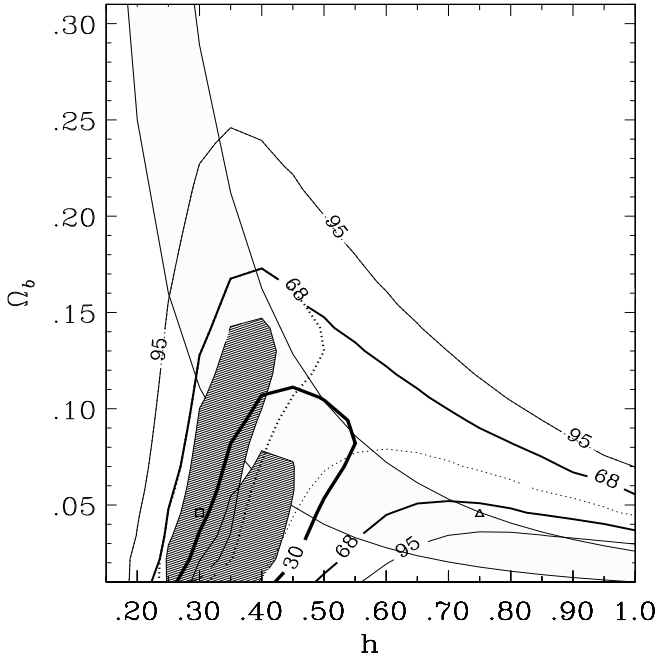,height=9cm,width=\hsize,bbllx=0pt,bblly=12pt,bburx=538pt,bbury=608pt}}
%
\caption{
Likelihood region and goodness-of-fit contours of $\chi^{2}$.
The similarity of this figure and Figure \protect\ref{fig:conv_a} is a 
measure of the domination of the fits by $A_{peak}$.
The solid and dotted lines are goodness-of-fit contours.
The solid contours (30\%, 68\%, 95\%) include 
our estimate of the Saskatoon calibration errors;
the dotted contours (68\%(thick), 95\%(thin)) do not.
The lower thin 95\% dotted contour is obvious but the upper 95\% dotted 
contour is identical to, and therefore hidden by, the solid 95\% contour.
The two dark grey areas are the $68\%$ likelihood regions for Sk$0$ (left)
and Sk$-14$ (right)(see text).
Small values of $h$ are preferred. 
The region preferred by big bang nucleosynthesis
($0.010 < \ob h^{2} < 0.026$) is the light grey band.
The small box and triangle at $\ob \approx 0.05$ are the two model spectra plotted 
in Figure \protect\ref{fig:data}.
} 
\label{fig:conv_chi}
\end{figure}

The CMB data can constrain any parameter that changes the
power spectrum at a level comparable with the error bars on the data.
Figure \ref{fig:conv_chi} displays the \chisq contours for models 
in the $h-\ob$ plane and contains one of the main results of this paper.
CMB observations alone exclude at $> 95\%$ CL, more than half of this parameter 
space.
The solid and dotted contours in Figure \ref{fig:conv_chi} are 
goodness-of-fit contours
labeled with the probabilities of finding
a \chisq less than the calculated value at that point.
 For example, in our case there are 21 degrees of freedom
(24 data points - 2 fitted parameters  - 1 COBE normalization).
The probability of obtaining a \chisq value less than 23.8  is 68.3 \%
assuming uncorrelated measurements and Gaussian errors.
Thus we have labeled the $\chi^{2} = 23.8$ contour `68'.
The levels plotted are $\chi^{2}= [17.2, 23.8, 32.6]$
corresponding to 30\%, 68\% and 95\% respectively.

The solid-line contours include our
estimate of the $14\%$ absolute calibration uncertainty shared by 
the 5 Saskatoon points.
The thick dotted (68\%) and thin dotted (95\%) contours 
do not include the Saskatoon calibration uncertainty.
Thus the Saskatoon errors do not affect the 95\% contour in the 
left and upper right, nor the left side of the 68\% contour.
The effect of the Saskatoon calibration uncertainties 
is discussed further in Section \ref{sec:skcal}.

Any data sensitive only to the peak height will yield 
contours as seen in Figure \ref{fig:conv_a}.
Figures \ref{fig:conv_a} and \ref{fig:conv_chi} are similar so peak height 
is dominating the fit.  If $A_{peak}$ were the only factor, then a region straddling the
$A_{peak} \sim 80\;\mu$K contour would be equally preferred by the data.
However the goodness-of-fit statistic prefers a region
that does not follow exactly any iso-$A_{peak}$ contour.
There is a preference for the low $h$ part of an iso-$A_{peak}$ region.

Lower $h$ is preferred because of the location of $\ell_{peak}$ in the data.
One can see in Figure \ref{fig:conv_cl} that at smaller $h$ the position of 
the peak is more aligned with the dotted (data) line.  
This can also be seen in Figure \ref{fig:conv_l} where the $\ell_{peak}$ values 
in the lower right are substantially less than the $\sim 280$ of the data.
This $\ell_{peak}$ mismatch disfavors
the high $h$ part of the preferred iso-$A_{peak}$ region.

\subsection{Saskatoon Calibration Error}
\label{sec:skcal}

The 5 Saskatoon (Sk) measurements (Netterfield \etal 1997)
apparently span the Doppler peak and play an important role in our
fitting procedure. All 5 points share a 14\% absolute calibration error
as indicated by the small squares above and below their central values
in Figure \ref{fig:data}.
There is no systematic way to handle systematic errors.
Adding 14\% errors in quadrature to the statistical errors is not
appropriate because these errors are 100\% correlated in the 5 Sk points.
We have examined the effect of these correlated 
errors is several ways. We have made \chisq surfaces from the CMB data:\\
\noindent $\bullet$ using the 5 Sk points as listed in Table 1 (``Sk$0$'')\\
\noindent $\bullet$ adding 14\% to the 5 Sk points, i.e., using the high small squares
in Figure \ref{fig:data} as the central values (``Sk$+14$'')\\
\noindent $\bullet$ subtracting 14\% from the 5 Sk points, i.e., using the lower small squares
in Figure \ref{fig:data} as the central values (``Sk$-14$'').\\
The minimum \chisq  for these three cases are
respectively, 20, 33 and 13  corresponding to 
goodness-of-fit contours of 48\%, 96\% and 9\%.
Thus Sk$0$ gives a reasonable fit, Sk$+14$ gives a bad fit and Sk$-14$ 
gives a fit that is a bit too good. 

We take a conservative approach in our treatment of these calibration errors.
We create a new surface by adopting at each point in the plane the 
minimum of these three \chisq  surfaces (Sk$+14$, Sk$0$ and Sk$-14$).
In Figure \ref{fig:conv_chi} we plot the iso-$\chi^{2}$ contours from this new surface.
We will call this the ``Sk$\pm14$'' case. 
For this new surface the Sk$-14$ \chisq surface determines the contours
in the lower right while the Sk$0$ \chisq surface determines the contours
to the left and upper right.
This is easy to understand with Figure \ref{fig:conv_a};
Sk$0$ has a higher peak than Sk$-14$. 
Thus Sk$0$ prefers $A_{peak} \sim 80 \mu$K and 
Sk$-14$ prefers $A_{peak} \sim 70\; \mu$K.
The Sk$+14$ surface is effectively eliminated  because of its poor goodness-of-fit.

The contours in Figure \ref{fig:conv_chi} are conservative in the sense
that they encompass a larger region than the corresponding contours
without including the Saskatoon calibration uncertainty.
The dotted contours are the 68\% and 95\% 
goodness-of-fit levels from Sk$0$. The solid lines are the
goodness-of-fit levels from our Sk$\pm 14\%$ method.
Thus Figure \ref{fig:conv_chi} shows where and by how much
the Saskatoon calibration uncertainty
reduces the constraining ability of the CMB data.

It is interesting to note that the rest of the data thinks that
the Saskatoon values are too high in the sense that the \chisq surface
from Sk$-14$ has a lower minimum than the Sk$0$ surface.
This can be quantified; we can guess at the correct
Sk calibration by using the non-Sk data and the models while 
treating the common calibration uncertainty of the Sk points as a free
parameter, i.e., we let the calibration vary and
assume that one of the models is correct.
The lowest \chisq is obtained if the Saskatoon points are reduced by 24\%;
the non-Sk data would like to see the Sk points all lowered by 24\%.
This preference is however shallow: $\chi^{2}_{min}(Sk-24\%) = 11.4$ while
$\chi^{2}_{min}(Sk-14\%) = 12.7$ and  $\chi^{2}_{min}(Sk-0\%)=19.9$.

The fact that the fit to Sk$-14$ is a bit too good can be interpreted as some combination of the
following:\\
\noindent $\bullet$ just chance\\
\noindent $\bullet$ non-Sk data prefer a peak height lower than the Sk measurements\\
\noindent $\bullet$ the error bars of the non-Sk points are over-estimated\\
\noindent $\bullet$ detections of small significance do not get reported or are reported as upper limits 
which have not been included in the data set\\
\noindent $\bullet$ observers are finding what they are supposed to find.\\

Reasonable goodness-of-fit is a prerequisite for \chisq minimum parameters that are meaningful.
In addition to the goodness-of-fit contours presented thus far we have used
the minima of the Sk$0$, Sk$-14$ and Sk$+14$ \chisq surfaces to define the 
best fit parameters and to define confidence levels around these minima.
This procedure is correct  when the goodness-of-fit is within some
plausible range, commonly  [5\%,95\%],
thus the Sk$+14$ minimum is of questionable value.
To obtain likelihood intervals on parameter values at the minimum
of a 2-D $\chi^{2}$ surface
one takes $\chi^{2}_{min} + 2.3$ as the 68\% contour
and $\chi^{2}_{min} + 6.2$ as the 95\% contour (e.g. Press \etal 1992).
In Figure \ref{fig:conv_chi}, the two dark grey areas are the $68\%$
likelihood regions for Sk$0$ (left) and Sk$-14$ (right). 
The preferred $h$ value is 
low and the confidence levels are smaller than the corresponding goodness-of-fit 
contours.
\begin{figure}[tbp]
\centerline{\psfig{figure=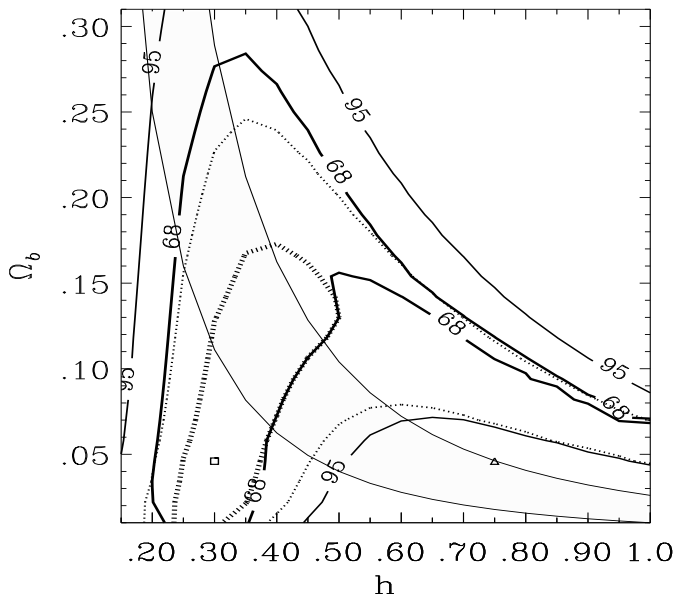,height=8cm,width=\hsize,bbllx=0pt,bblly=12pt,bburx=538pt,bbury=608pt}}
%
\caption{
Same as Figure \protect\ref{fig:conv_chi} except instead of the
Saskatoon uncertainty, here we show the effect of the COBE 
normalization uncertainty.
The dotted contours are the same as in Figure \protect\ref{fig:conv_chi}.
The solid contours include our estimate of the effect of the 
COBE normalization uncertainty.
They are the contours of a \chisq surface constructed from the minima
of three \chisq surfaces, one each for $Q=16.4$, $18.0$ and $19.6$.
The higher $Q$ value lets $\ob$ be as high as 0.28.
The limits on $h$ do not change substantially and thus
small values of $h$ are still preferred independent of the 
COBE normalization uncertainty.
} 
\label{fig:conv_q}
\end{figure}

\subsection{Normalization Uncertainty}

So far we have assumed the COBE normalization 
$Q|_{n_{s}=1} = 18\; \mu$K at
$\ell = 6$.
However there is a $1.6\; \mu$K overall uncertainty on this value.
To include this uncertainty in our results we make \chisq contours for 
$Q = 16.4$, $18.0$ and $19.6\;\mu$K.
%
The minimum of the surfaces (produced by a procedure analogous to that used to make 
Figure \ref{fig:conv_chi}) is shown in Figure \ref{fig:conv_q} where the dotted contours
do not, and the solid contours do, include the COBE normalization  uncertainty.
The normalization uncertainty
increases the upper limit on $\ob$ from 0.17 to 0.28 (68\% CL).
The limits on $h$ are robust to the normalization 
uncertainty. The lower right is still excluded.

\subsection{Sensitivity to ``Outliers''}

One possible danger in using \chisq goodness-of-fit contours is the influence of outliers.
%
The overall level (goodness-of-fit) 
and possibly the shape of  the contours around
the minimum can be controlled by outliers.
This depends on whether the models differ much at the $\ell_{eff}$ of the 
suspected outlier. For example in Figure \ref{fig:data}, the COBE point at 
$\ell=25$ is an ``outlier'' which does nothing more than raise the entire \chisq 
surface, since all models are the same at $\ell \sim 25$.
This is not necessarily the case for the low MAX point (MAX HR)
and certainly not for the 5 Saskatoon points. 
When the MAX HR point is excluded from the Sk$0$ surface calculation, 
the goodness-of-fit of the minimum improves from
$\chi_{min}^{2} = 20$ to $\chi^{2}_{min}= 15$. 
The size and shape of the resulting
contours do not change significantly, probably because
at $\ell \approx 140$ the models are not substantially different.
\begin{figure}[htbp]
\centerline{\psfig{figure=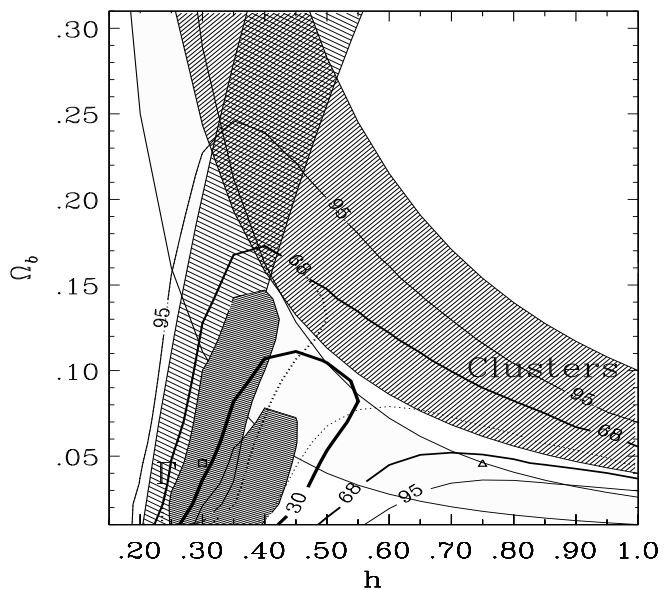,height=8cm,width=\hsize,bbllx=0pt,bblly=12pt,bburx=538pt,bbury=608pt}}
%
\caption{
Same as Figure \protect\ref{fig:conv_chi} except we have added two 
more constraints: 
Clusters: $0.04 < \ob h^{3/2}  < 0.10$ (White \etal 1993) and
the galaxy and cluster power spectrum shape parameter: 
$0.20 < \Gamma  < 0.30$ (Peacock \& Dodds 1994).
} 
\label{fig:conv_other}
\end{figure}

\subsection{Result Summary}

Our main results for the $h-\ob$ plane are summarized in 
Figure \ref{fig:conv_chi}.
In the context of flat CDM universes, CMB observations can exclude more than 
half of the $h - \ob$ parameter space explored.
The CMB data favor low values of Hubble's constant; $h \sim 0.35$. 
A higher value such as $h=0.75$ is permitted by the goodness-of-fit
contours if $0.05 < \ob < 0.09$
but these values are not within the 1-$\sigma$ likelihood region
and are excluded by big bang nucleosynthesis (BBN).
Low values of $\ob$ are preferred ($\ob \sim 0.03$) but the \chisq 
minimum is shallow and we obtain $\ob < 0.28$ (68\% CL).

These limits include our estimates for the Sk absolute calibration
errors and the COBE normalization uncertainty.
They are conservative in the sense that 
they are larger than the corresponding error bars around the parameters at the 
minimum of the \chisq surfaces.
Outliers do not seem to be biasing these results.

The CMB and BBN are independent observations.
Thus, the overlapping of regions preferred by BBN and 
the CMB did not necessarily have to exist.
The fact that they do is a consistency test for these two pillars of
the big bang scenario.

The uncertainties in the BBN limits are dominated by systematic errors.
Therefore we have used `weak' BBN limits to try to avoid over-constraining the 
parameters. We have increased the limits of Copi \etal (1995), $0.010 < \ob h^{2}  < 0.020$, 
to include the higher values obtained by Tytler \& Burles (1997), $0.022 < \ob h^{2}  < 0.026$.
The credibility of the BBN constraints used here is important because their
incompatibility with the CMB constraints at $h \ga 70$ is what excludes these models.
If one takes the maximum upper limit on $\ob h^{2}$ as $0.045$ (Reeves 1994) then
at the 68\% CL all values of $h$ between $0.30$ and $1.0$ are permitted by even the
combination of the CMB and BBN limits.

Figure \ref{fig:conv_other} is the same as Figure \ref{fig:conv_chi} 
except we have added two more constraints:
the White \etal (1993) limits from cluster baryon fractions:
$0.04 < \ob h^{3/2}  < 0.10$, and
the CDM shape parameter from a synthesis of the power spectra of galaxies and clusters: 
$0.20 < \Gamma  < 0.30$ (Peacock \& Dodds 1994).
The similarity of the $\Gamma$ limits and the Sk$0$ contours is interesting.
The inconsistency of BBN limits with the White \etal (1993) limits (assuming $\oo = 1$)
is sometimes invoked as an argument for $\oo < 1$. 
However, as Figure \ref{fig:conv_other} shows, this inconsistency disappears for the low
$h$ values favored by the CMB data.
An $\oo=1$ model with $h\approx 0.40$, $\ob \approx 0.15$ and  $\oc \approx 0.85$
is permitted by constraints from the CMB data, BBN, cluster baryon fractions 
and $\Gamma$.

\begin{figure}[htbp]
\centerline{\psfig{figure=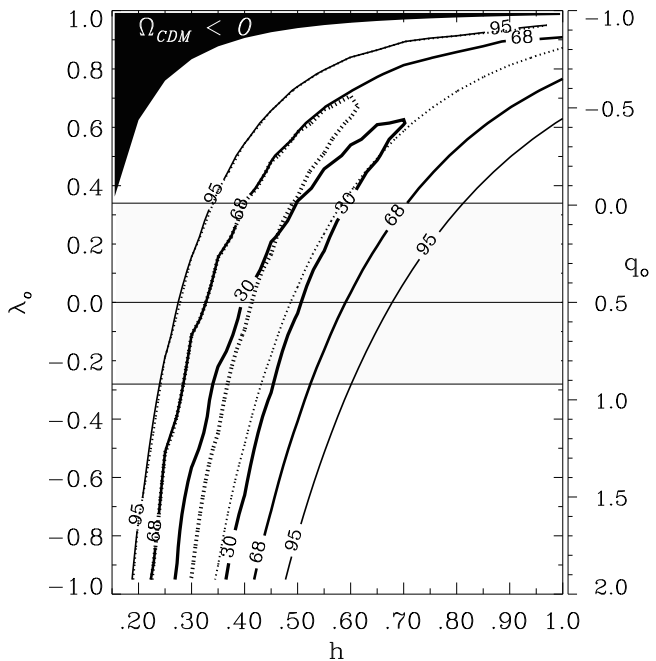,height=9cm,width=\hsize,bbllx=18pt,bblly=40pt,bburx=594pt,bbury=630pt}}
%
\caption{
Goodness-of-fit contours of $\chi^{2}$ with
$\ob h^{2} = 0.015$.
As in Figure \protect\ref{fig:conv_chi} the solid contours include the Saskatoon
calibration uncertainty, the dotted do not.
The grey area marks the permitted area (1 $\sigma$) from new supernovae results 
(Perlmutter \etal 1997).
For all models in this plot, $\oo + \lambda_{o} = 1$.
} 
\label{fig:hlam_chi}
\end{figure}
\begin{figure}[htbp]
\centerline{\psfig{figure=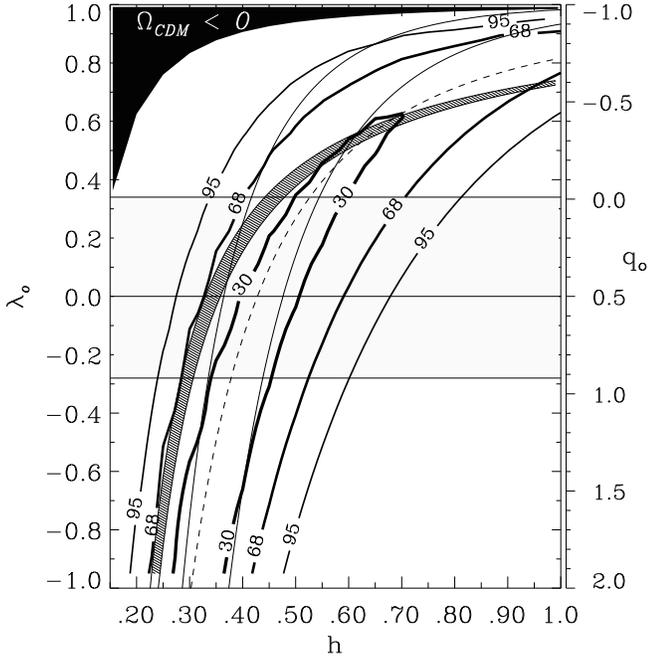,height=9cm,width=\hsize,bbllx=18pt,bblly=40pt,bburx=594pt,bbury=630pt}}
%
\caption{
Same as Figure \protect\ref{fig:hlam_chi} except we have added two
more constraints and left out the dotted contours for legibility.
Constraints on the age of the universe ($13.7 < t_{o} < 17.9$)
from the oldest stars (Bolte \& Hogan 1995) are indicated by the thin solid lines.
The dark grey area marks limits on the shape parameter:  
$0.20 < \Gamma  < 0.30$ (Peacock \& Dodds 1994).
The thick dashed line through the minimum is $\oc h^{2} = (0.41)^{2}$.
} 
\label{fig:hlam_other}
\end{figure}
\section{ $h - \ol$ Results and Discussion}

In the previous section we described results in the $h-\ob$ plane
($k=0$, $\oo = 1$, $\ol = 0$).
In this section we present results for the $h-\ol$ plane.
($k=0$, $\oo + \ol = 1$).
Figure \ref{fig:hlam_chi} presents the \chisq contours from fitting
the data to these models.
We are exploring a plane orthogonal to Figure \ref{fig:conv_chi}.
The same procedure used to make the contours in Figure \ref{fig:conv_chi} was used here.
The notation is also the same; the solid contours include the Saskatoon
calibration uncertainties, the dotted do not.

Figures \ref{fig:hlam_a} and \ref{fig:hlam_chi} are similar 
for the same reason that Figures \ref{fig:conv_a} and 
\ref{fig:conv_chi} are similar:
$A_{peak}$ is the dominant feature of the fit.
There is a slight preference for the low $h$ side of the preferred 
iso-$A_{peak}$ contour.
Figure \ref{fig:hlam_l} helps explain this preference.
For a given $\ol$, $\ell_{peak} \sim 280$ is at a lower $h$ than
the $A_{peak} \sim 80$ contour.

In this plane we obtain $0.23 < h < 0.72$ where these limits 
include our estimates for the uncertainties from the Saskatoon calibration,
the COBE normalization and the BBN interval $0.010 < \ob h^{2} < 0.026$.
Assuming $\ob h^{2} = 0.015$, the CMB data yield $\ol \la 0.9$ (68\% CL) with lower values 
preferred. This is weaker than the traditional dynamical limit $\ol \la 0.8$.
A standard flat-$\ol$ model is $\oo \sim 0.3$ and $\ol \sim 0.7$
with $h\approx .70$
This model is fully consistent with the CMB data.
However the new SNIa results (Perlmutter \etal 1997) rule it out;
$-0.28 \la \ol \la 0.34$ ($1 \sigma$). 


\section{ Summary and Discussion}
We have explored solutions to the 
Boltzmann equation in two 2-D parameter spaces
within the context of COBE normalized, Gaussian, adiabatic initial
conditions in spatially flat universes with Harrison-Zel'dovich temperature fluctuations.
We have presented the topology which controls the shapes of the acceptable regions for the 
cosmological parameters $h$, $\ob$, and $\ol$.
We have fit a compilation of the most recent CMB 
measurements to the models and have identified regions
favored and excluded by the data, i.e.,
we have fit the Boltzmann solutions to the CMB data 
and obtained constraints on the cosmological parameters $h$, $\ob$, and $\ol$.

Figures \ref{fig:conv_chi} and \ref{fig:hlam_chi} contain the main results 
of this paper and are plots of the goodness-of-fit contours of the 
\chisq from equation (1) using the data in Table 1.
The solid contours mark the regions preferred by the data and include
the Saskatoon calibration uncertainty.
We obtain the following results:\\
$\bullet$ Recent CMB data are precise enough to prefer distinct regions
of parameter space and rule out most of the $h-\ob$ and $h-\ol$ planes at 
the 95\% CL.\\
$\bullet$ The CMB data favor low values of Hubble's constant; $h \approx 0.35$. 
A higher value such as $h=0.75$ is permitted by the goodness-of-fit
contours if $0.05 < \ob < 0.09$
but these values are not within the 1-$\sigma$ likelihood region
and are excluded by big bang nucleosynthesis (BBN).\\
%
$\bullet$ Low values of $\ob$ are preferred ($\ob \sim 0.03$) but the \chisq 
minimum is shallow and we obtain $\ob < 0.28$.\\
$\bullet$ The CMB regions overlap with BBN.
The fact that they do is a consistency test for these two pillars of
the big bang scenario.\\ 
%
$\bullet$ A $\oo=1$ model with $h\approx 0.40$, $\ob \approx 0.15$ and  
$\oc \approx 0.85$ is permitted by constraints from the CMB data, 
BBN, cluster baryon fractions and the shape parameter $\Gamma$ derived 
from the mass density power spectra of galaxies and clusters.\\ 
$\bullet$ For flat-$\ol$ models, the CMB data, combined with BBN  constraints 
exclude most of the $h - \ol$ plane.\\ 
$\bullet$ Models with $\oo \approx  0.3$, $\ol \approx 0.7$ with $h \approx 0.75$ 
are fully consistent with the CMB data but are excluded by the strict new 
$q_{o}$ limits from supernovae.\\
$\bullet$ A combination of CMB data goodness-of-fit statistics, BBN and supernovae 
constraints in the $h-\ol$ plane,  limits Hubble's constant to the 
interval $0.23 < h < 0.72$.

In this analysis we have made estimates of the errors associated with
the Saskatoon absolute calibration uncertainty, the COBE normalization
uncertainty, the uncertainty in BBN and we have considered the influence
of possible outliers.
We have assumed Gaussian adiabatic initial conditions.
We have conditioned on the values $n_{s}=1$ and $Y_{He}=0.24$.
We have ignored the possiblity of open universes, tilted spectra, early reionization
and any gravity wave contribution to the spectra.

Our $h$ result is `low' and inconsistent with several other recent $h$ measurements 
($0.65 \la h \la 0.80$) in the sense that our 68\% goodness-of-fit contours do not, but 
our 95\% contours do include these higher $h$ values. 
The theoretical advantages of a low Hubble constant have been presented in Bartlett \etal (1995).
For example, if $\Omega_{o} = 1$ then a Hubble constant of $0.70$ implies an age of 9.3 Gyr, younger
than the estimated ages of many globular clusters.
However $h\sim 0.40$ yields  an age of the universe of 16 Gyr, much more in accord with globular 
cluster ages.

If new data show that $A_{peak}\sim 70\;\mu$K rather than $80\;\mu$K and 
$\ell_{peak} \sim 230$ rather than 280, then the range $0.65 \la h \la 0.80$
may be acceptable to both the CMB and BBN constraints.

\subsection{Acknowledgements}
We gratefully acknowledge the use of the Boltzmann code kindly
provided by Uros Seljak and Matias Zaldarriaga. We thank Martin White, Douglas Scott and the MAX group for help
in assembling the required experimental window functions.
We thank our referee, Douglas Scott, for comments that improved the clarity of this
paper.
We benefited from statistical discussions with Jean-Marie Hameury and
Olivier Bienaym\'e.
C.H.L. acknowledges
support from the French Minist\`ere des Affaires Etrang\`eres. 
D.B. is supported by the Praxis XXI CIENCIA-BD/2790/93 grant attributed by JNICT, Portugal.

\clearpage
{\scriptsize
\begin{table}[t]
\begin{center}
\caption{Data Used in the \chisq Fit$^{a}$ and Plotted in Figure \protect\ref{fig:data}}
\vspace{2pt}
\begin{tabular}{|l|l|r|c|} \hline         
Experiment & reference  &$\ell_{eff}$& $\delta T_{\ell_{eff}}^{data} \pm \sigma ^{data}(\mu$K$)^{b}$\\
\hline
DMR1    &\protect\cite{hin96}&  3&$ 27.9_{-  4.0}^{+  5.6}$\\
DMR2    &\protect\cite{hin96}&  7&$ 24.6_{-  2.8}^{+  3.6}$\\
DMR3    &\protect\cite{hin96}& 14&$ 30.8_{-  3.1}^{+  3.4}$\\
DMR4    &\protect\cite{hin96}& 25&$  1.2_{-  1.2}^{+ 14.4}$\\
FIRS    &\protect\cite{gan94}& 10&$ 29.4_{-  7.7}^{+  7.8}$\\
Tenerife&\protect\cite{ha96a}& 19&$ 34.1_{- 12.5}^{+ 12.5}$\\
SP91    &\protect\cite{gun95}& 60&$ 30.2_{-  5.5}^{+  8.9}$\\
SP94    &\protect\cite{gun95}& 60&$ 36.3_{-  6.1}^{+ 13.6}$\\
Pyth1   &\protect\cite{pla96}& 91&$ 54.0_{- 12.0}^{+ 14.0}$\\
Pyth2   &\protect\cite{pla96}&176&$ 58.0_{- 13.0}^{+ 15.0}$\\
ARGO1   &\protect\cite{deb94}& 95&$ 39.1_{-  8.7}^{+  8.7}$\\
ARGO2   &\protect\cite{mas96}& 95&$ 46.8_{- 12.1}^{+  9.5}$\\
MAX GUM &\protect\cite{tan96}&138&$ 54.5_{- 10.9}^{+ 16.4}$\\
MAX ID  &\protect\cite{tan96}&138&$ 46.3_{- 13.6}^{+ 21.8}$\\
MAX SH  &\protect\cite{tan96}&138&$ 49.1_{- 16.4}^{+ 21.8}$\\
MAX HR  &\protect\cite{tan96}&138&$ 32.7_{-  8.2}^{+ 10.9}$\\
MAX PH  &\protect\cite{tan96}&138&$ 51.8_{- 10.9}^{+ 19.1}$\\
Sk1     &\protect\cite{net97}& 86&$ 49.0_{-  5.0}^{+  8.0}$\\
Sk2     &\protect\cite{net97}&166&$ 69.0_{-  6.0}^{+  7.0}$\\
Sk3     &\protect\cite{net97}&236&$ 85.0_{-  8.0}^{+ 10.0}$\\
Sk4     &\protect\cite{net97}&285&$ 86.0_{- 10.0}^{+ 12.0}$\\
Sk5     &\protect\cite{net97}&348&$ 69.0_{- 28.0}^{+ 19.0}$\\
CAT1    &\protect\cite{sco96}&396&$ 51.8_{- 13.6}^{+ 13.6}$\\
CAT2    &\protect\cite{sco96}&607&$ 49.1_{- 13.7}^{+ 19.1}$\\
\hline
\end{tabular}
\end{center}
$^{a}$ 
CMB anisotropy detections reported in publications in 1994 and later.
Many of the new result papers include reanalyses of older results.
MSAM is not included because of substantial spatial and filter function
overlap with Saskatoon.\\

\noindent $^{b}\:$$Q_{flat}= Q_{rms-PS}|_{n_{s}=1}= (\frac{5}{12})^{1/2}\:\delta T_{\ell_{eff}}^{data}$.

\end{table}
}   
\end{document}